# Comment: Lancaster Probabilities and Gibbs Sampling

Gérard Letac

## 1. LANCASTER PROBABILITIES AS THE PROPER FRAMEWORK

It is a pleasure to congratulate the authors for this excellent, original and pedagogical paper. I read a preliminary draft at the end of 2006 and I then mentioned to the authors that their work should be set within the framework of Lancaster probabilities, a remoted corner of the theory of probability, now described in their Section 6.1. The reader is referred to Lancaster (1958, 1963, 1975) and the synthesis by Koudou (1995, 1996) for more details.

Given probabilities $\mu(dx)$ and $\nu(dy)$ on spaces $\mathcal{X}$ and $\mathcal{Y}$, and given orthonormal bases $p = (p_n(x))$ and $q = (q_n(y))$ of $L^2(\mu)$ and $L^2(\nu)$, a probability $\sigma$ on $\mathcal{X} \times \mathcal{Y}$ is said to be of the Lancaster type if either there exists a sequence $\rho = (\rho_n)$ in $\ell^2$ such that

$$\sigma(dx, dy) = \left[\sum_n \rho_n p_n(x) q_n(y)\right] \mu(dx) \nu(dy)$$

or $\sigma$ is a weak limit of such probabilities. Alternatively, one can say that the sequence of signed measures $[\sum_{n=0}^N \rho_n p_n(x) q_n(y)] \mu(dx)\nu(dy)$ converges weakly toward the probability $\sigma$ when $N \to \infty$ (here $\rho$ does not need to be in $\ell^2$). An acceptable sequence $\rho = (\rho_n)$ is called a Lancaster sequence for the quadruple $(\mu, \nu, p, q)$. If $p_0 = q_0 = 1$ the margins of $\sigma$ are $(\mu, \nu)$. Writing

$$\sigma(dx, dy) = \mu(dx) K(x, dy) = \nu(dy) L(y, dx)$$

the probability kernel of the "$x$-chain" considered in the paper is

$$k(x, dx') = \int_\mathcal{Y} K(x, dy) L(y, dx')$$

G. Letac is Professor, Laboratoire de Statistique et Probabilités, Université Paul Sabatier, 31062 Toulouse, France e-mail: letac@cict.fr.

$$= \left[\sum_n \rho_n^2 p_n(x) p_n(x')\right] \mu(dx')$$

which clearly shows that $p_n$ is an eigenfunction for the eigenvalue $\rho_n^2$ of the operator $Tf(x) = \int_\mathcal{X} f(x') \cdot k(x, dx')$.

I will not comment here on the multivariate case $\mathcal{X} = \mathbb{R}^k$ and $\mathcal{Y} = \mathbb{R}^m$. Everything which is known about Lancaster probabilities and which is specific to this case is mentioned in Section 7 of the paper. To my knowledge, the Lancaster probabilities on the torus $(\mathbb{R}/\mathbb{Z})^2$ associated to the trigonometric orthonormal polynomials have never been considered.

For the present time, the richest case is obviously the one where $\mathcal{X} = \mathcal{Y} = \mathbb{R}$ and where $p = (p_n)$ and $q = (q_n)$ are the orthonormal polynomials obtained by the Schmidt orthonormalization process in $L^2(\mu)$ and $L^2(\nu)$ applied to the sequences $(x^n)$ and $(y^n)$, assuming furthermore that $\int e^{a|x|} \mu(dx)$ and $\int e^{a|y|} \nu(dy)$ are finite for some $a > 0$. In the sequel, the term "Lancaster probabilities" will refer only to this real case. The following should be specified clearly:

> Saying that conditions H1, H2 and H3 of Section 3 are all fulfilled is equivalent to saying that $P(dx, d\theta)$ is a Lancaster probability.

An elegant example can be found in Buja (1990, page 1049) with

$$\sigma(dx, dy) = \frac{a+b}{B(a, b)} x^{a-1} y^{b-1} \mathbf{1}_A(x, y) \, dx \, dy$$

where $a, b > 0$ and $A = \{(x, y); \ 0 < x, y; \ x + y < 1\}$. The margins are $\mu(dx) = \beta_{a, b+1}(dx)$ and $\nu(dy) = \beta_{b, a+1}(dy)$ and the Lancaster sequence is

$$\rho_n = \frac{(-1)^n \sqrt{ab}}{\sqrt{(a+n)(b+n)}}.$$

The present paper on discussion is based on three observations. The first one is crucial: the two-components Gibbs sampler is very easy to perform with a Lancaster probability. This is the statement in Theorem 3.1. Parts a and b are well known but part c is elegant and surprising.

## 2. NATURAL EXPONENTIAL FAMILIES

In order to explain the other two observations, let us introduce some notation: a (not necessarily bounded) positive measure $\mu$ on $\mathbb{R}$ is said to be in $\mathcal{M}(\mathbb{R})$ if it is not concentrated on one point and if its Laplace transform

$$L_\mu(\theta) = e^{k_\mu(\theta)} = \int_{-\infty}^{\infty} e^{\theta x} \mu(dx)$$

is such that the interior $\Theta(\mu)$ of the interval $D(\mu) = \{\theta \in \mathbb{R};\ L_\mu(\theta) < \infty\}$ is not empty. To such a $\mu \in \mathcal{M}(\mathbb{R})$ one associates the one-dimensional natural exponential family (NEF):

$$\begin{aligned}F &= F(\mu) \\ &= \{P(\mu,\theta)(dx) = e^{\theta x - k_\mu(\theta)}\mu(dx); \theta \in \Theta(\mu)\}.\end{aligned}$$

Since $\theta \mapsto k_\mu(\theta)$ is strictly convex on $\Theta(\mu)$ the map $\theta \mapsto m = k'(\theta) = \int_{-\infty}^{\infty} x P(\mu,\theta)(dx)$ is injective and its image $M_F = k'_\mu(\Theta(\mu))$ is an open interval called the domain of the means of $F$. One denotes by $m \mapsto \theta = \psi_\mu(m)$ the inverse map from $M_F$ to $\Theta(\mu)$. Finally we say $F$ or $\mu$ is steep if $M_F$ is the interior of the convex support of $\mu$. For instance $D(\mu) = \Theta(\mu)$ (in this case $F$ is said to be regular) implies that $F$ is steep. The converse is not true. Diaconis and Ylvisaker (1979) show that if $F$ is regular, if $x_0 \in M_F$ and if $\lambda > 0$ then there exists a constant $C(x_0, \lambda)$ such that

$$\pi_{x_0,\lambda}(d\theta) = C(x_0,\lambda) e^{\lambda(\theta x_0 - k_\mu(\theta))} \mathbf{1}_{\Theta(\mu)}(\theta)\,d\theta$$

is a probability. We call $\{\pi_{x_0,\lambda};\ x_0 \in M_F,\ \lambda > 0\}$ the Diaconis–Ylvisaker family associated to the NEF $F$. We now reparameterize it by the mean. More specifically, denote by

$$\nu_{x_0,\lambda}(dm) = C(x_0,\lambda)\exp\lambda(x_0\psi_\mu(m) - k_\mu(\psi_\mu(m)))$$
$$\cdot \psi'_\mu(m)\mathbf{1}_{M_F}(m)\,dm$$

the image of $\pi_{x_0,\lambda}(d\theta)$ by $\theta \mapsto m = k'_\mu(\theta)$. Finally consider the distribution on $\mathbb{R}^2$ defined by

$$\sigma(dx, dm) = P(\mu, \psi_\mu(m))(dx)\nu_{x_0,\lambda}(dm).$$

Note that the marginal distribution $\mu_1(dx)$ of $\sigma(dx, dm)$ does not belong to $F$ except in the normal case. (Proving this is an amusing exercise. It even holds when the reference measure $d\theta$ in the Diaconis–Ylvisaker family is replaced by any other positive measure.[1]) The second observation of the paper, and a quite original one, is that $\sigma(dx, dm)$ is a Lancaster probability if $F$ is either binomial (Section 4.1), or Poisson (Section 4.2), or Gaussian (Section 4.3). An element of the Diaconis–Ylvisaker family associated with the binomial case $B(\theta, n)$ is the beta distribution $\nu_1(d\theta) = \beta_{a,b}(d\theta)$ and the marginal distribution of $X$ is the so-called hypergeometric distribution

(1) $$\mu_1(dx) = \sum_{k=0}^{n} \binom{n}{k} \frac{(a)_k (b)_{n-k}}{(a+b)_n} \delta_k(dx).$$

The construction of a Lancaster probability with these margins $(\mu_1, \nu_1)$ have never been done before. Here the Lancaster sequence is $\rho_j = n!/(a+b+n)_j(n-j)!$ for $0 \leq j \leq n$ and $\rho_j = 0$ if $n < j$. The Lancaster probabilities obtained for $F$ = Poisson and $F$ = Gaussian are familiar and are mentioned in Koudou (1996, Section 3.3) and studied in Koudou (1995).

My guess is that these 3 types of NEF are the only ones with such a property: this is obviously false for the three other quadratic NEF (Negative binomial, gamma, hyperbolic), for which $\nu_{x_0,\lambda}(dm)$ has very few moments. The reader can check for example that the same is true for the NEF generated by a stable law of parameter $\alpha \in (0,1)$ concentrated on $(0,\infty)$ and defined by $k_\mu(\theta) = -c(-\theta)^\alpha$: recall that $\alpha = 1/2$ gives the celebrated Inverse Gaussian distributions (the case $\alpha \in [1,2)$ has not to be investigated since it is not steep).

In order to explain the content of the third observation of the paper, we introduce the Jorgensen set $\Lambda(\mu)$ of $\mu \in \mathcal{M}(\mathbb{R})$. It is the set of $\lambda \geq 0$ such that for $\lambda > 0$ there exists $\mu_\lambda \in \mathcal{M}(\mathbb{R})$ such that $\Theta(\mu_\lambda) = \Theta(\mu)$ and such that $L_{\mu_\lambda} = (L_\mu)^\lambda$. We impose $0 \in \Lambda(\mu)$. For instance $\Lambda(\mu) = [0, \infty)$ if and only if $F(\mu)$ is made of infinitely divisible distributions. On the other hand $\Lambda(\mu)$ is the set of nonnegative integers if $\mu = \delta_0 + \delta_1$, namely if $F$ is the Bernoulli family. In general $\Lambda(\mu)$ can be a quite complicated additive semigroup: see Letac, Malouche and Maurer (2002) for its description when $\mu$ is the convolution of a negative binomial distribution with a

---

[1] The family $G$ obtained in this way is also a conjugate family to $F$, which means that the a posteriori distribution $\pi(d\theta|x)$ is in $G$ when the a priori distribution $\pi$ is in $G$. For this reason we do speak of the Diaconis–Ylvisaker family instead of the conjugate family of the paper, even if the later has the characteristic property mentioned in Section 2.3.2.

Bernoulli distribution. Now consider $\mu \in M(\mathbb{R})$ and $\lambda$ and $\eta$ in $\Lambda(\mu)$. Let

$$(X, Y) \sim P(\mu_\lambda, \theta) \otimes P(\mu_\eta, \theta).$$

Write $S = X + Y \sim P(\mu_{\lambda+\eta}, \theta)$ (the distribution of $Y$ knowing $S$ does not depend on $\theta$) and denote by $\sigma(ds, dy)$ the joint distribution of $(S, Y)$. The authors observe that, when $F$ happens to be a quadratic NEF, $\sigma$ is a Lancaster probability: this is the essence of Section 5. However, this is a particular case of the following classical result mentioned in Eagleson (1964): suppose that $\lambda, \eta, \xi$ are in $\Lambda(\mu)$ and let

$$(X, Y, Z) \sim P(\mu_\lambda, \theta) \otimes P(\mu_\eta, \theta) \otimes P(\mu_\xi, \theta).$$

Denote by $\sigma(ds, dt)$ the joint distribution of $(S, T) = (X + Y, Y + Z)$. Then $\sigma$ is a Lancaster probability if $F$ is a quadratic NEF. More specifically if $(p_n^{(\lambda)})$ is the sequence of the orthonormal polynomials for $P(\mu_\lambda, \theta)$ and if $1/c_n(\lambda)$ is the positive square root of the coefficient of $x^n$ in $p_n^{(\lambda)}$ the corresponding Lancaster sequence is

$$(2) \qquad \rho_n = \frac{c_n(\eta)}{\sqrt{c_n(\lambda + \eta) c_n(\eta + \xi)}}.$$

Thus Section 5 is based on the particular case $\lambda = n_1, \eta = n_2, \xi = 0$ of this result.

## 3. FINDING ALL LANCASTER FAMILIES WITH GIVEN MARGINS

Given a pair of probabilities $(\mu, \nu)$ on $\mathbb{R}$ such that $\int e^{a|x|} \mu(dx)$ and $\int e^{a|y|} \nu(dy)$ are finite for some $a > 0$, consider the set $L(\mu, \nu)$ of Lancaster probabilities $\sigma$ with margins $(\mu, \nu)$ and the set $S(\mu, \nu)$ of corresponding Lancaster sequences $\rho = (\rho_n)_{n=0}^\infty$. They are isomorphic compact convex sets which are completely known if we know their extreme points. We denote by $I(\mu)$ the smallest closed interval $I$ such that $\mu(I) = 1$. We consider several cases:

Case A. $I(\mu)$ is bounded, $I(\nu)$ is unbounded.
Case B. $I(\mu) = \mathbb{R}$ and $I(\nu)$ is a half-line.
Case C. $I(\mu) = I(\nu) = \mathbb{R}$.
Case D. $I(\mu)$ and $I(\nu)$ are half-lines.
Case E. $I(\mu)$ and $I(\nu)$ are bounded.

Cases A and B are easy: the only Lancaster probability is the product measure. Denote by $a_n > 0$ and by $b_n > 0$ the coefficients of $x^n$ in the orthonormal polynomials $p_n$ and $q_n$. Case C is quite interesting: from a remarkable result of Tyan and Thomas (1975), extending an idea of Sarmanov and Bratoeva (1967), which says that if $\gamma = \liminf(a_{2n}/b_{2n})^{1/2n}$ and if $\rho \in S(\mu, \nu)$, there exists a probability $\alpha(dt)$ on $[-\gamma, \gamma]$ such that $a_n \rho_n/b_n = \int_{-\gamma}^{\gamma} t^n \alpha(dt)$. Similarly in the case D, assuming without loss of generality that $I(\mu)$ and $I(\nu)$ are positive half-lines and if $\gamma = \liminf(a_n/b_n)^{1/n}$ then there exists a probability $\alpha(dt)$ on $[0, \gamma]$ such that $a_n \rho_n/b_n = \int_0^\gamma t^n \alpha(dt)$. The results of Tyan and Thomas (1975) can also essentially be found again in Tyan, Derin and Thomas (1976) and have been rediscovered by Christian Berg, quoted in Ismail (2005, page 114) who does not seem to be aware of this previous work.

We shall speak about case E later on. Note that for $\mu = \nu$ the results by Tyan and Thomas are quite exciting since they mean that a Lancaster sequence must be the moment sequence of a probability either on $[-1, 1]$ (case C) or on $[0, 1]$ (case D). If we are fortunate enough to prove that $\rho_n = t^n$ is a Lancaster sequence for all $t \in [-1, 1]$ (case C) or all $t \in [0, 1]$, by the theorems of Tyan and Thomas, we have a complete description of the Lancaster probabilities $L(\mu, \mu)$ since they are parameterized by the probabilities $\alpha$ on $[-1, 1]$ or on $[0, 1]$. Interestingly enough, this is known to happen only for 4 types of $\mu$: Gaussian, Poisson, negative binomial and gamma. The corresponding Lancaster probabilities (see Bar-Lev et al., 1994) are the only ones which belong to a two-dimensional natural exponential family with variance function of the form

$$\begin{bmatrix} a(m_1) & f(m_1, m_2) \\ f(m_1, m_2) & a(m_2) \end{bmatrix}.$$

More specifically one can conjecture the following:

- If $I(\mu) = \mathbb{R}$ and if $(t^n)$ is in $S(\mu, \mu)$ for all $t \in [-1, 1]$ then $\mu$ is Gaussian.
- If $I(\mu) = [0, \infty)$ and if $(t^n)$ is in $S(\mu, \mu)$ for all $t \in [0, 1]$ then $\mu$ is gamma, or Poisson, or negative binomial.

In the gamma case, it is interesting to consider the classical two-dimensional distribution of Kibble (1941) and Moran (1967) with correlation $r \in [0, 1]$ and Jorgensen parameter $q$. It can be defined by its Laplace transform

$$\int_0^\infty \int_0^\infty e^{-sx-ty} \sigma_r(dx, dy)$$
$$= (1 + s + t + (1-r)st)^{-q}.$$

As observed by D'jachenko (1962), this probability is actually a Lancaster probability with $\rho_n = r^n$, and thus an extremal one (the last three references are taken from Johnson and Kotz, 1972, pages 479–482). This means that in general $\sigma$ is a Lancaster probability for the gamma margins $\mu = \nu = \gamma_q$ if and only if it is a mixing of Kibble and Moran distributions, which means that there exists a probability distribution $\alpha(dr)$ on $[0,1]$ such that

$$\int_0^\infty \int_0^\infty e^{-sx-ty}\sigma(dx,dy)$$
$$= \int_0^1 (1+s+t+(1-r)st)^{-q}\alpha(dr).$$

Take for instance $\alpha(dr) = \beta_{\eta,q-\eta}(dr)$ to get back (2) for the gamma case and $\lambda = \xi = q - \eta$.

For the cases C and D and for $\nu$ not an affine transformation of $\mu$, there is no known example where the set of the extreme points of $L(\mu,\nu)$ can be completely described. Koudou (1995, 1996) has shown that $\rho_n = t^n$ is a Lancaster sequence:

- for $\mu = P_a$ and $\nu = P_b$ ($P_a$ means Poisson distribution with mean $a$) for $0 \le t \le (a/b)^{1/2}$ if $a \le b$;
- for $\mu = NB_{a,\lambda}$ and $\nu = NB_{a,\lambda}$ [the negative binomial distribution $NB_{a,\lambda}$ is $(1-a)^\lambda \sum_{n=0}^\infty \frac{(\lambda)_n}{n!} a^n \cdot \delta_n(dx)$] for $0 \le t \le (a/b)^{1/2}$ if $a \le b$;
- for $\mu = NB_{a,\lambda}$ and $\nu = \gamma_\lambda$ for $0 \le t \le a^{1/2}$.

In these three cases, one can conjecture that one has obtained all the extreme points of $S(\mu,\nu)$.

Consider a hyperbolic distribution $\mu_q$ as described in Section 2.4 and simply defined by $L_{\mu_q}(\theta) = (\cos\theta)^{-q}$ with $q > 0$ and $\Theta(\mu_q) = (-\frac{\pi}{2}, \frac{\pi}{2})$. Lai and Vere-Jones (1975) have proved that $(t^n)$ is never in $S(\mu_q, \mu_q)$ (an other proof is in Bar-Lev et al., 1994). Formula (2) applies here with $c_n(q) = \frac{(q)_n}{n!}$. In (2) we take $0 \le \eta \le q$ and $\lambda = \xi = q - \eta$ to show that the sequence

$$\rho_n = \frac{c_n(\eta)}{c_n(q)} = \frac{1}{B(\eta, q-\eta)} \int_0^1 t^n t^{\eta-1}(1-t)^{q-\eta-1}\, dt$$

is an element of $S(\mu_q, \mu_q)$. This illustrates the results of Tyan and Thomas with $\alpha(dt) = \beta_{\eta,q-\eta}(dt)$. One can conjecture (as done by Lai and Vere-Jones for $q = 1$) that such a Lancaster sequence indexed by $\eta \in [0,q]$ is an extreme point of $S(\mu_q, \mu_q)$, and that all extreme points are of this type.

## 4. THE CASE WHERE $\mu$ AND $\nu$ HAVE BOUNDED SUPPORT

This is the case E above. For the variety of results already obtained in the literature, this is the richest case. For future research, it is the most challenging. If $\mu = \nu$ suppose that there exists $x_0$ such that $|p_n(x)| \le p_n(x_0)$ $\mu$ almost surely, and consider

$$(3) \quad K(x,y,z) = \sum_{n=0}^\infty \frac{1}{p_n(x_0)} p_n(x)p_n(y)p_n(z).$$

Koudou (1995) has shown that $K \ge 0$ for almost all $(x,y,z)$ in the $\mu$ sense implies that the extreme points of $S(\mu,\mu)$ are defined by $\rho_n = p_n(x)/p_n(x_0)$ when $x$ describes the support of $\mu$. This extends a remarkable paper by Eagleson (1969) devoted to the case where $\mu$ is discrete with finite support, where it is shown in particular that $K \ge 0$ when $\mu$ is a binomial distribution. As mentioned in the paper, the analysis by Koudou (1996) of Gasper's (1971) delicate results shows that $K \ge 0$ when $\mu = \beta_{a,b}$ is a beta distribution such that $a, b \ge 1/2$ [note that the case $\min(a,b) < 1/2$ is open].

The particular case $a = b \ge 1/2$ deserves a special mention. Using the transformation $x \mapsto 2x - 1$, we first move the distributions from $[0,1]$ to $[-1,1]$ and we introduce

$$\Delta(x,y,z) = 1 - x^2 - y^2 - z^2 + 2xyz.$$

For $-1 < z < 1$ we consider the plane domain $U_z = \{(x,y);\ \Delta > 0\}$. This domain is limited by an ellipse $E_z$ tangent to the sides of the unit square $[-1,1]^2$. Denote $\mu_a(dx) = \frac{2^{1-2a}}{B(a,a)}(1-x^2)^{a-1}\mathbf{1}_{(-1,1)}(x)\,dx$. The number $x_0$ involved in the definition of $K$ in (3) is 1, and the polynomials $p_n$ are the Jacobi polynomials with suitable parameters and normalized such that they become orthonormal with respect to $\mu_\alpha$. With these notation, $K$ is zero outside of $U_z$ and is equal to

$$K_a(x,y,z)$$
$$= C(\alpha)[(1-x^2)(1-y^2)(1-z^2)]^{1-a}\Delta^{a-3/2}$$

in $U_z$. The important point is the following. For $z \in (-1,1)$ consider the extremal Lancaster probabilities $\sigma_z(dx,dy) = K_a(x,y,z)\mu_a(dx)\mu_a(dy)$. These Lancaster probabilities $\sigma_z$ are the only ones (together with the centered nonsingular Gaussian distributions with covariance of the type $\begin{bmatrix} a & b \\ b & a \end{bmatrix}$) to be elliptically contoured. More specifically, let $E = \mathbb{R}^2$ have the Euclidean structure such that $U_z$ is the unit disk. Saying that $\sigma_z$ is elliptically contoured means that $\sigma_z$ is invariant by the orthogonal group $\mathbb{O}(E)$ of this Euclidean structure. This characterization is the consequence of an elegant result of McGraw and Wagner (1968). While most of the "results" about elliptically contoured distributions in $\mathbb{R}^d$ are trivially reduced to considerations about rotational invariant distributions, this is not the case here. The reason is that the canonical basis of $\mathbb{R}^2$ is structurally important for Lancaster probabilities. In the other hand this canonical basis is not orthonormal for the Euclidean structure associated with a given elliptically contoured distribution and this makes attractive the McGraw and Wagner result.

Koudou (1995) shows that we have $K \geq 0$ when

$$\mu(dx) = \frac{q+1}{2\pi} \frac{\sqrt{p^2 - x^2}}{1 - x^2} \mathbf{1}_{(-p,p)}(x)\, dx$$

where $q > 0$ and $p = 2\sqrt{q}/(1+q)$. When $q$ is an integer this strange probability is the Plancherel measure of the Gelfand pair associated to the homogeneous tree where every vertex has $q + 1$ neighbors. The corresponding polynomials are called the Cartier–Dunau polynomials in the literature (see Arnaud, 1994). A general theory of the probabilities $\mu$ with bounded support such that the function $K$ of (3) is positive could be a subject of research. As an example, I do not know whether $K \geq 0$ or not when $\mu$ is the hypergeometric distribution (1) considered in the paper, where the orthonormal polynomials are the Hahn polynomials.

When $\mu$ and $\nu$ are two probabilities with bounded support such that $\nu$ is not an affine transform of $\mu$, the search of extreme points of the Lancaster measures does not seem to have been done for any example. Suppose that we have found some $\rho \in S(\mu, \nu)$. A good way to create other elements of $S(\mu, \nu)$ is to pick $a \in S(\mu, \mu)$ and $b \in S(\nu, \nu)$. It is easy to see that $(a_n \rho_n b_n)_{n=0}^{\infty}$ is also in $S(\mu, \nu)$. Applying this remark to the interesting pair $(\mu_1, \nu_1)$ defined by (1) and to the new Lancaster sequence $\rho_j = n!/(a+b+n)_j(n-j)!$ discovered by the authors would lead to a better understanding of $S(\mu_1, \nu_1)$.

## 5. CONCLUSION

We referred to several bright papers by Eagleson, Koudou, McGraw and Wagner or Tyan and Thomas, and to a genuine masterpiece by Gasper. Many stimulating questions and conjectures remain, regarding in particular special functions and group theory through the function $K$. The present paper shows us how unexpectedly these bivariate probabilities can be important for very practical questions: it will be in turn a new landmark of the theory of Lancaster probabilities.

## REFERENCES

Arnaud, J.-P. (1994). Stationary processes indexed by a homogeneous tree. *Ann. Probab.* **22** 195–218. MR1258874

Bar-Lev, S., Bshouty, D., Enis, P., Letac, G., Lu, I.-L. and Richards, D. (1994). The diagonal multivariate natural exponential families and their classification. *J. Theoret. Probab.* **7** 883–929. MR1295545

Buja, A. C. (1990). Remarks on functional canonical variates, alternating least square methods and ACE. *Ann. Statist.* **18** 1032–1069. MR1062698

Diaconis, P. and Ylvisaker, D. (1979). Conjugate priors for exponential families. *Ann. Statist.* **7** 269–281. MR0520238

D'jachenko, Z. N. (1962). On a form of bivariate $\gamma$-distribution. *Nauchnye Trudy Leningradskoi Lesotekhnicheskoi Akademii* **94** 5–17. (In Russian.)

Eagleson, G. (1964). Polynomial expansions of bivariate distributions. *Ann. Math. Statist.* **35** 1208–1215. MR0168055

Eagleson, G. K. (1969). A characterization theorem for positive definite sequences on the Krawtchouk polynomials. *Aust. J. Statist.* **21** 256–265. MR0328162

Gasper, G. (1971). Banach algebra for Jacobi series and positivity of a kernel. *Ann. Math.* **95** 261–280. MR0310536

Ismail, M. (2005). *Classical and Quantum Orthogonal Polynomials*. Cambridge Univ. Press. MR2191786

Johnson, N. L. and Kotz, S. (1972). *Continuous Multivariate Distributions*. Wiley, New York. MR0418337

Kibble, S. W. (1941). A two-variate gamma type distribution. *Sankhyā* **5** 137–150. MR0007218

Koudou, A. E. (1995). Problèmes de marges et familles exponentielles naturelles. Thèse, Université Paul Sabatier, Toulouse.

Koudou, A. E. (1996). Probabilities de Lancaster. *Exp. Math.* **14** 247–275. MR1409004

Lai, C. D. and Vere-Jones, D. (1975). Odd man out. The Meixner hypergeometric distribution. *Aust. J. Statist.* **21** 256–265. MR0561951

Lancaster, H. O. (1958). The structure of bivariate distributions. *Ann. Math. Statist.* **29** 719–736. MR0102150

Lancaster, H. O. (1963). Correlations and canonical forms of bivariate distributions. *Ann. Math. Statist.* **34** 434–443. MR0146912

Lancaster, H. O. (1975). Joint probability distributions in the Meixner classes. *J. Roy. Statist. Soc. Ser. B* **37** 532–538. MR0394971

Letac, G., Malouche, D. and Maurer, S. (2002). The real powers of the convolution of a negative binomial and a Bernoulli distribution. *Proc. Amer. Math. Soc.* **130** 2107–2114. MR1896047

Moran, P. A. P. (1967). Testing for correlation between non-negative variates. *Biometrika* **54** 385–394. MR0221711

McGraw, D. K. and Wagner, J. G. (1968). Elliptically symmetric distributions. *IEEE Trans. Inform. Theory* **14** 110–120.

Sarmanov, O. V. and Bratoeva, Z. N. (1967). Probabilistic properties of bilinear expansions of Hermite polynomials. *Theory Probab. Appl.* **12** 470–481. MR0216541

Tyan, S. and Thomas, J. B. (1975). Characterization of a class of bivariate distribution functions. *J. Multivariate Anal.* **5** 227–235. MR0375625

Tyan, S., Derin, H. and Thomas, J. B. (1976). Two necessary conditions on the representations of bivariate distributions by polynomials. *Ann. Statist.* **4** 216–222. MR0391384